%% file: main_enase.tex
\documentclass[a4paper,twoside]{article}

\usepackage{epsfig}
\usepackage{subcaption}
\usepackage{calc}
\usepackage{amssymb}
\usepackage{amstext}
\usepackage{amsmath}
\usepackage{amsthm}
\usepackage{multicol}
\usepackage{pslatex}
\usepackage{apalike}
\usepackage{algorithm2e}
\usepackage[bottom]{footmisc}
\usepackage{SCITEPRESS}     % Please add other packages that you may need BEFORE the SCITEPRESS.sty package.

\usepackage{url}
\usepackage{rotating,adjustbox,multirow,multicol}
\usepackage[most]{tcolorbox}

\usepackage{ulem}
\usepackage{comment}
\usepackage{xcolor}

\begin{document}

\title{Variability-Driven User-Story Generation \\using LLM and Triadic Concept Analysis}

%\title{Give me the user-story set for my next new system:\\ Leveraging Variability of a Family of User-story Sets \\using LLM and Triadic Concept Analysis}

\author{\authorname{Alexandre Bazin \sup{1},
%\orcidAuthor{0000-0002-4268-8961}, 
Alain Gutierrez \sup{1},
Marianne Huchard \sup{1},
%\orcidAuthor{0000-0002-6309-7503},
\\Pierre Martin \sup{2,3},
%\orcidAuthor{0000-0002-4874-5795},
and Yulin (Huaxi) Zhang \sup{4}
%\orcidAuthor{0009-0008-3019-245X}
}
\affiliation{\sup{1}LIRMM, Univ. Montpellier, CNRS, Montpellier, France}
\affiliation{\sup{2}CIRAD, UPR AIDA, F-34398 Montpellier France}
\affiliation{\sup{3}AIDA, Univ. Montpellier, CIRAD, Montpellier, France}
\affiliation{\sup{4}EPROAD, Université de Picardie Jules Verne, 
Amiens, France}
\email{alexandre.bazin@lirmm.fr,
alain.gutierrez@lirmm.fr,
marianne.huchard@lirmm.fr,
\\pierre.martin@cirad.fr,
yulin.zhang@u-picardie.fr}
}


\abstract{A widely used Agile practice for requirements is to produce a set of user stories (also called ``agile product backlog''), which roughly includes a list of pairs (role, feature), where the role handles the feature for a certain purpose. In the context of Software Product Lines, the requirements for a family of similar systems is thus a family of user-story sets, one per system, leading to a 3-dimensional dataset composed of sets of triples (system, role, feature). In this paper, we combine Triadic Concept Analysis (TCA) and Large Language Model (LLM) prompting to suggest the user-story set required to develop a new system relying on the variability logic of an existing system family. This process consists in 1) computing 3-dimensional variability expressed as a set of TCA implications, 2) providing the designer with intelligible design options, 3) capturing the designer's selection of options, 4) proposing a first user-story set corresponding to this selection, 5) consolidating its validity according to the implications identified in step 1, while completing it if necessary, and 6) leveraging LLM 
% -CS- commonsense knowledge 
to have a more comprehensive website. This process is evaluated with a dataset comprising the user-story sets of 67 similar-purpose websites.}

\onecolumn \maketitle \normalsize \setcounter{footnote}{0} \vfill

\input{sections/introduction}
\input{sections/approach}

\input{sections/materialsmethods}

\input{sections/findings}

\input{sections/relatedWork}

\input{sections/conclusion}

\section*{\uppercase{Acknowledgements}}
This work was supported by the ANR SmartFCA project, 
%\footnote{\url{https://www.smartfca.org/}}, 
Grant ANR-21-CE23-0023 of the French National Research Agency.

\bibliographystyle{apalike}
{\small
\bibliography{biblioshort}}

\end{document}

%% file: sections/introduction.tex
\section{\uppercase{Introduction}}
\label{sec:introduction}

At the requirements stage, a widespread practice of the Agile paradigm is to provide a set of user-stories (also called ``agile product backlog''), where a user-story is a brief sentence expressing the fact that a 'persona' (or role) wants to perform an 'action' (or have access to a feature) with a certain 'purpose' \cite{UsAsRequirment}.
In the context of Software Product Lines (SPL, \cite{DBLPbooks/daglib/0015277}), the requirements for a family of similar systems are therefore a family of similar user-story sets, one set per system. User-story sets are 
usually 
stored 
to support product line requirements documentation, guide the development, and are connected to the source code.

In this paper, we address the issue of \textit{building the user-story set for a new system based on the variability logic of an existing system family, according to design options provided by the system designer.}
We investigate 
a process that combines Triadic Concept Analysis (TCA) \cite{DBLP:conf/iccs/LehmannW95} and Large Language Model (LLM) prompting with the system designer input to suggest the user-story set for the new system. 
The design options of the new system are provided at an intermediate level of description (e.g. e-commerce), rather than at the feature level (e.g. pay), to alleviate the configuration work.

Our approach operates at the two stages of the traditional SPL framework \cite{DBLPbooks/daglib/0015277}. 
It contributes to the \textit{domain engineering} stage by building a variability model for requirements which is composed of (1) a set of triadic implications \cite{DBLP:conf/iccs/GanterO04} and (2) an intelligible set of design options provided by LLM.
At the \textit{application engineering} stage, a selection of design options, made by the software designer, leads LLM to propose  an initial user-story set. Then LLM uses the triadic implication set to consolidate the validity of the proposed user-story set, completing it if necessary to get a nearly valid configuration. Finally, we leverage LLM to propose user-stories related to the current user-story set, in order to have an even more comprehensive website.

The process is evaluated through a case study using a dataset of the literature \cite{DBLP:journals/ijar/BazinGHMT24}. Results are encouraging and indicate that the combination of the rigor of TCA and 
knowledge brought by LLM would be beneficial. This dataset is composed of user-story sets of 67 similar websites in several domains (mangas and derived products, martial art equipment, board games and video games).

%%%%%% Outline

Section \ref{sec:approach} presents Triadic Concept Analysis (TCA) and  the complexity for a software designer to leverage such outputs. 
Section \ref{sec:casestudy} presents the approach. It  outlines the process and presents the material and method adopted to address the case study. 
Section \ref{sec:findings} presents and discusses the results.
Related work is presented in Section  \ref{sec:relatedwork}, and we conclude in Section  \ref{sec:conclusion} with a summary and future work.

%% file: sections/approach.tex
\section{\uppercase{Triadic Concept Analysis}}
\label{sec:approach}

\paragraph{TCA in a nutshell}
\label{sec:triadic}

Formal Concept Analysis (FCA)~\cite{DBLP:books/daglib/0095956} is a mathematical framework that aims at structuring  information found in data in the form of binary relations. It starts with a binary relation where  objects are described by attributes  (see Table~\ref{tab:context}).

\begin{table}[htb]
\caption{A  relation between systems as objects 
%($MyManga$, $MangaStore$, $MangaHome$) 
and features as attributes, 
%($search$, $view~comment$, $manage~cart$)
inspired from \cite{DBLP:journals/ijar/BazinGHMT24}.}
\label{tab:context}
%\resizebox{\textwidth}{!}{
\centering
\adjustbox{width=0.7\columnwidth}{
\begin{tabular}{c|ccc}
 & $search$ & $view~comment$ & $manage~cart$  \\
% & $s$ & $vc$ & $mc$  \\
 \hline
 $MyManga$ & $\times$ &  &  $\times$ \\
 $MangaStore$ & $\times$ & &   \\
 $MangaHome$ & & $\times$ & $\times$   \\
 \hline
\end{tabular}
}
\end{table}

In binary relations, an implication is a pattern of the form $A\rightarrow B$ where $A$ and $B$ are attribute sets such that every object described by the attributes of $A$ is also described by the attributes of $B$. For instance, the implication $\{view~comment\}\rightarrow\{manage~cart\}$ holds in Table~\ref{tab:context} as all the systems offering the $view~comment$ feature (only $MangaHome$) also offer the $manage~cart$ feature.

The user-stories we consider are ternary relations between systems, roles and features (see Table~\ref{tab:3context}).
%\footnote{In this paper, we do not consider the purpose, as done in the used dataset of the literature.}. 
TCA~\cite{DBLP:conf/iccs/LehmannW95} has been developed in order to exploit the more complex information they contain.
In Table~\ref{tab:3context}, a final user can search in all systems, and view comments only in $MangaHome$.
A product manager can manage cart in $MyManga$ and $MangaHome$, and view comments in $MangaHome$.

\begin{table}[htb]
    \caption{A ternary relation between systems ($MyManga$, $MangaStore$, $MangaHome$), features (search $s$, view~comment $vc$, manage~cart $mc$) and roles ($FinalUser$, $Administrator$, $ProductManager$) \cite{DBLP:journals/ijar/BazinGHMT24}.}
    \label{tab:3context}
\resizebox{\columnwidth}{!}{
    \begin{tabular}{c|ccc||ccc||ccc}
     & $s$ & $vc$ & $mc$ & $s$ & $vc$ & $mc$ & $s$ & $vc$ & $mc$ \\
     \hline
    ~ $MyManga$~ & $\times$ & & & $\times$ & & $\times$ & & & $\times$ \\
    ~ $MangaStore$~ & $\times$ & & & $\times$ & & & & & \\
    ~ $MangaHome$~ & $\times$ & $\times$ & & & $\times$ & & & $\times$ & $\times$ \\
     \hline
     \multicolumn{1}{c|}{} & \multicolumn{3}{c||}{$FinalUser$} & \multicolumn{3}{c||}{$Administrator$} &
     \multicolumn{3}{c}{$ProductManager$} 
\end{tabular}
}
\end{table}

Implications in a triadic setting are more diversified than in their dyadic counterpart~\cite{DBLP:conf/iccs/GanterO04}. Indeed, one can be interested in implications between features, between roles, or between the allocations of specific features to specific roles, \textit{i.e.} pairs (feature,role) or symmetrically pairs (role,feature). To obtain these latter rules, triadic data are brought back to a dyadic view: a binary relation is created by taking the Cartesian product of the required dimension as attributes and the Cartesian product of the other dimensions as objects. For instance, Table~\ref{tab:pfr} depicts a binary relation between systems and the pairs (feature,role) they offer. The implication $\{(s,A)\}\rightarrow\{(s,FU)\}$ holds and means that all the systems that offer the $search$ feature to administrators also offer it to final users. Two systems ($MyManga$ and $MangaStore$) offer the implication premise $(s,A)$. This number is called the \textit{support} of the implication. 

\begin{table}[htb]
\begin{center}
\caption{\label{tab:pfr} A binary relation between systems and pairs composed of a feature (search $s$, view comment $vc$, manage comment $mc$) and a role (FinalUser $FU$, Administrator $A$, ProductManager $PM$).}
\resizebox{\columnwidth}{!}{
    \begin{tabular}{c|ccccccccc}
         & ($s$,$FU$) & ($vc$,$FU$) & ($mc$,$FU$) & ($s$,$A$) & ($vc$,$A$) & ($mc$,$A$) & ($s$,$PM$) & ($vc$,$PM$) & ($mc$,$PM$)  \\
         \hline
        $MyManga$ & $\times$ & & & $\times$ & & $\times$ & & &  \\
        $MangaStore$ & $\times$ & & & $\times$ & & & & &  \\
        $MangaHome$ & $\times$ & $\times$ & & & $\times$ & & & $\times$ & $\times$  \\
    \end{tabular}
    }
\end{center}
\end{table}

In this paper, we use only implications between pairs (feature,role), and whose premise is a singleton, to prevent LLM from facing excessive computation challenges in this first study. 

\paragraph{Handling implications to design a new system}
\label{sec:difficulty}

Let's illustrate the limit of handling implications to design a new system using a small dataset taken from \cite{DBLP:journals/ijar/BazinGHMT24}, which  introduces sets of user-stories from four manga-related websites.
%%%%%
Two %examples of
implications between pairs (role, feature) are shown below:
\begin{tcolorbox}[breakable, enhanced]
{\footnotesize
\begin{verbatim}
<4>  => (user;search)
<2> (communityManager;moderateComment) 
    => (user;viewComment)
(...)
\end{verbatim}
}
\end{tcolorbox}

%%%%%
In this set, an implication is expressed as $<n>(r1;f1) => (r2;f2)$, where $n$ is the support that informs on the number of websites which provide the premise $(r1;f1)$ of the implication. 
Such an implication thus means that in $n$ websites, when role $r1$ can perform feature $f1$ (premise), then role $r2$ can perform feature $f2$ (conclusion).

\begin{figure*}[htb]
  \includegraphics[width=\textwidth]{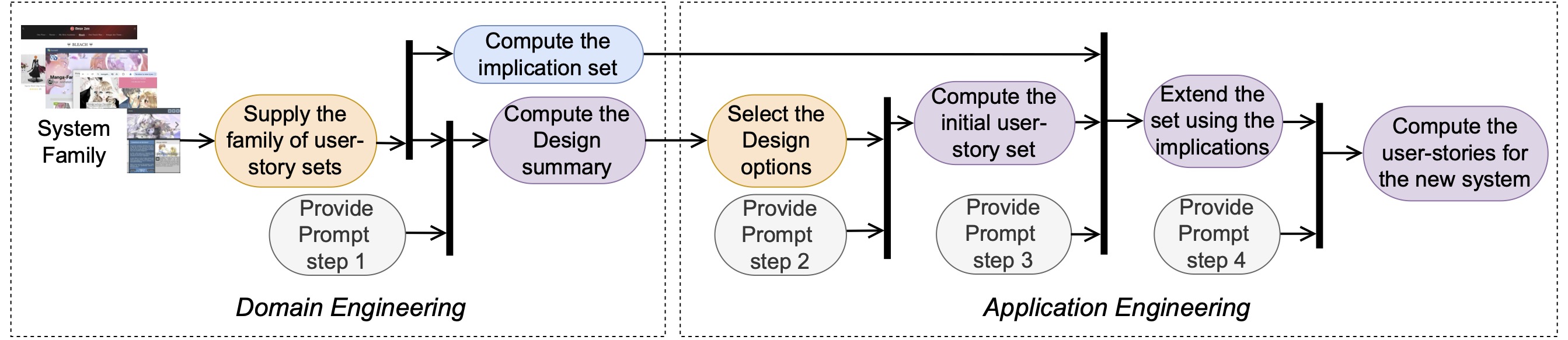}
  \caption{Overview of the process within the software product line framework. 
  In this UML activity diagram, a color informs on the actuator:
   orange, violet, blue, and grey refer respectively to website designer, LLM, TCA, and prompt designer.}
  \label{fig:overview}
\end{figure*}

User story sets and binary implications capture a large part of the variability logic of the Manga-related website family. 
They fix the vocabulary on role names (e.g.  'registeredUser' corresponds to 'premium user', 'subscriber', etc.), and feature names (e.g. 'CRUDproducts' corresponds to 'manageProductsDB').
The implications indicate, for instance \textit{all websites provide search to users} (1st implication, held by all 4 systems); \textit{when a community manager can moderate comments, then users can view the comments} (2nd implication, held by 2 systems). 

Implications being numerous, information is difficult to grasp by a software designer. Nor is it usable, as it does not give a synthetic report of the high-level options available (such as e-commerce or community management) and the logical dependencies to be respected when developing a new website.
This is where LLM comes into play, with its ability to summarize and leverage 
knowledge to recommend features and dependencies in a more general setting.

%% file: sections/materialsmethods.tex
\section{\uppercase{Approach}}
\label{sec:casestudy}

%This section outlines the process, introduces the software and dataset used to conduct the case study,  the 4 steps prompt and the addressed research questions.

\paragraph{Process overview}
\label{sec:schema}

Figure \ref{fig:overview} outlines the process 
%proposed in this paper 
within the SPL framework \cite{DBLPbooks/daglib/0015277}.
The first activities take place at the \textit{Domain engineering} stage, which focuses, in this work, on identifying commonalities and variability in the requirements of the systems provided as input.
In a first step, the family of user-story sets is extracted from the system family storage, and then communicated with a prompt (Prompt step 1) to LLM. In return, the latter computes the main design options and provides a \textit{design summary}.
In parallel, the family of user-story sets is parsed with TCA, producing a set of implications that express logical dependencies between user-stories.

A second group of activities takes place in the \textit{Application engineering} stage, which aims to produce a set of user-stories for the new system to be developed.
As a second step of the process, the designers \textit{select the design options} in the list proposed in the Design summary, then LLM provides the initial user-story set corresponding to this selection using Prompt step 2.
Prompt step 3 asks LLM to extend the initial user-story set using the TCA implications.
Finally, using Prompt step 4, LLM is requested to  add or remove user-stories to get a more comprehensive system.
All files of the case study are available online\footnote{ \url{https://doi.org/10.18167/DVN1/GNJMAV/}}.  

%\begin{figure*}[htb]
%  \includegraphics[width=\textwidth]{figures_sources/overview4.jpg}
%  \caption{Overview of the process within the software product line framework. 
%  In this UML activity diagram, a color informs on the actuator:
%   orange, violet, blue, and grey refer respectively to website designer, LLM, TCA, and prompt designer.}
%  \label{fig:overview}
%\end{figure*}

\paragraph{Tool and Dataset}
\label{sec:dataset}

The LLM adopted to conduct the case study is the general ChatGPT  4.0 model, to benefit from its latest enhancements. The purpose is to allow anyone to use or reproduce our results. 
Regarding the dataset, we used the file $ALL\_System\_Role\_ActingVerb.csv$ reported in \cite{DBLP:journals/ijar/BazinGHMT24}.
This dataset gathers the sets of user-stories extracted in 2023 by students from 67 similar-purpose  websites within the domains of mangas and derived products, sport equipment for martial arts, board games, and video games. The extraction has been supervised and the result has been reviewed and standardized.
This dataset contains 1546 triples $(system, role, action~verb)$, where 67 systems, 17 roles and 30 action~verbs are involved, giving 91 user-stories. From this dataset, TCA computed 
687  implications for the relation  $system \times (role;feature)$ given in the companion repository\footnote{\url{https://doi.org/10.18167/DVN1/BWCC71}}.

\paragraph{Prompt design}
\label{sec:prompting}

To write the prompt, we have complied with the recommendations of OpenAI\footnote{\url{https://platform.openai.com/docs/guides/prompt-engineering}} and %the 
literature \cite{schulhoff2024promptreportsystematicsurvey,DBLP:conf/msr/MondalBR24,DBLP:journals/corr/abs-2302-11382}.
(i) A \textit{role} (persona) is assigned to LLM in order to clarify its position in relation to the tasks it has to perform. 
(ii) A \textit{context} is given as a key part of the prompt to guide LLM towards a relevant outcome. This context outlines the aim and the framework required by LLM to understand the data and the  tasks.
(iii) The chat is \textit{decomposed} into several tasks.
(iv) LLM is asked to \textit{review} its answers. This often helps LLM correct mistakes and achieve a better result.
(v) The \textit{syntax} is explained and illustrated, notably when it is complex, such as implications, i.e. a mathematical logic formulation preceded by the implication support. Clearly identifying the syntax information on the prompt greatly improves the results.
Finally, (vi) using \textit{tags} allows to differentiate different parts within the prompt (e.g. \texttt{<Context>}, \texttt{<Syntax of data>}), and to refer data and results (e.g. \texttt{<User story data>}, \texttt{<Result step2>}).

The final prompt and the associated answers are presented below in a representative chat. 
As the aim of this  %paragraph 
section is to present the prompt, long answers are cut short and the removed parts are replaced by an ellipsis \textit{(...)}.
To facilitate data transmission to LLM, two files are uploaded in the chat, i.e. $UserStoriesInitial.csv$ and $ImplicationsInitial.txt$ that contain respectively the user-story set family for step 1 and the implication set computed by TCA for step 3.

%Prompt step 1, presented below, asks LLM for a summary of high-level design options. 
Prompt step 1 asks LLM to provide a summary of high-level design options. This prompt is presented below.

\begin{tcolorbox}[breakable, enhanced]
{\footnotesize
Step 1: 

$<$Role$>$: Your role is to assist software engineers in designing websites using the principles of software product lines. 
(...)

$<$Context$>$: (...) you will be guided by a provided description of various similar websites in the targeted business domain (Manga websites) to identify the high-level design options. 
(...)

$<$Syntax of data$>$: Each line in the raw data represents a user story of a website, expressed as a triple: “website, role, feature.” 

$<$User story Data$>$: UserStoriesInitial.csv
	
---\\
$<$Task 1$>$: Based on $<$User story Data$>$, can you categorize these role-features into logical groups that represent distinct website feature modules
(...)

$<$Task 2$>$: Please review the groups to ensure accuracy, 
(...)

}
\end{tcolorbox}

An answer excerpt to Prompt step 1, is as follows:
\begin{tcolorbox}[breakable, enhanced]
{\footnotesize
(...)
Proposed Feature Modules:\\
\textit{(...)}\\
    \textbf{Payment \& Subscription Management}\\
        Roles: Finaluser, Subscriber, FinancialManager
        Features: pay, subscribe, unsubscribe, cancel\\
\textit{(...)} \\  \textbf{Promotions \& Marketing}\\
        Roles: MarketingManager, SocialMediaInfluencer
        Features: promote, advertise, campaign, evaluate
 \textit{(...)}  
 
 }
\end{tcolorbox}

Prompt step 2 is then completed by the selected Design options. 

\begin{tcolorbox}[breakable, enhanced]
{\footnotesize
Step 2: 

$<$Context$>$: In this step 2, you will identify the set of needed (role;feature) pairs to implement a set of chosen high-level design options (logical groups).

---\\
$<$Task 1$>$: Based on result of Task 2 of Step 1, can you list the (role;feature) pairs (i.e. the user-stories) that are needed to implement (...)
[\textbf{Payment \& Subscription Management and Promotions \& Marketing}].

$<$Task 2$>$: Review your result.

$<$Formatted result step 2$>$ Show the plain list of pairs (role;feature) from these groups (...)

}
\end{tcolorbox}

In the answer, as presented  below, LLM lists the relevant user-stories for the selected options.

\begin{tcolorbox}[breakable, enhanced]
{\footnotesize
(Finaluser;pay)\\
%(Finaluser;subscribe)\\
%(Finaluser;unsubscribe)\\
(Financialmanager;validate)\\
(Intermediateseller;subscribe)
\textit{(...)}

}
\end{tcolorbox}

In Prompt step 3, LLM is asked to consider the TCA implications in order to have a user-story set respecting the dataset logical dependencies:  

\begin{tcolorbox}[breakable, enhanced]
{\footnotesize
Step 3: 

$<$Context$>$: In this step, you will leverage logical dependencies (...)
Using them, you will recommend an initial set of user stories 
(...)

$<$Binary Implications$>$: 

In the general case, a binary implication follows the format “$<$n$>$ (r1;f1) $=>$ (r2;f2) ”, where (...)

$<$Implications derived for Input Data$>$: ImplicationsInitial.txt

---\\
$<$Task 1$>$: The result of step 2 ($<$Formatted result step 2$>$) lists the pairs (role;feature) corresponding to the user stories associated to the high-level options chosen by the software engineer. 
Can you apply the $<$Implications derived for input data$>$, to recommend the set of user stories (pairs (r;f)) that describe this website? (...)

$<$Task 2$>$: Review your result

$<$Result 1 step 3$>$ Show the plain list of applied implications (...)

$<$Result 2 step 3$>$ Show the final plain list of (role;feature) pairs (i.e. user stories). (...)

}
\end{tcolorbox}

In its answer (see below), LLM lists the applied implications, and updates the user-story list.

\begin{tcolorbox}[breakable, enhanced]
{\footnotesize
Result 1 step 3 (Applied Implications):

(Finaluser;unsubscribe) $=>$ (Finaluser;add)

(Finaluser;makeanoffer) $=>$ (Finaluser;add)
\textit{(...)}

Result 2 step 3 (Final Set of (Role;Feature) Pairs):

(Administrator;login)
(Administrator;manage)
(Contentcreator;add)
\textit{(...)}
  
}
\end{tcolorbox}

%\newpage
Prompt step 4 asks LLM to enlarge the proposal:

\begin{tcolorbox}[breakable, enhanced]
{\footnotesize
Step 4: 

$<$Context$>$ In this step, we would like to refine the previous result 
(...)

$<$Task$>$ 
(...)
would you add or remove any (role; feature) pairs in $<$Result 2 step 3$>$?
(...)

$<$Result step 4$>$ Updated list of user-stories (...)

}
\end{tcolorbox}

The final answer of LLM completes the website design, with explanations:

\begin{tcolorbox}[breakable, enhanced]
{\footnotesize
(...)
Here are the adjustments made:

    Additions:\\
        (Finaluser;download): Users might want the ability to download Manga chapters or entire volumes for offline reading, (...)\\
    \textit{(...)}

    Removals:
        None; (..)

Here’s the final set of user stories for a comprehensive Manga website:

Result Step 4:

Administrator;login\\
Administrator;manage\\
\textit{(...)}
   
}
\end{tcolorbox}

\paragraph{Investigated questions / Aim of the evaluation}

The case study aims to assess LLM ability:
to produce a relevant set of user-stories for a new system in the SPL framework, and 
to combine its knowledge with logical dependencies extracted from existing systems using a logic-based method, i.e. TCA.
We thus focused on these main research questions:
\begin{itemize}
\item (Q1) Is LLM able to properly summarize the shared or specific high-level design options of the existing system family?
\item (Q2) Is LLM able to leverage logical dependencies to derive a nearly valid set of user-stories, starting from user-stories selected by the software engineer? 
\item (Q3) Is LLM able to extend the set of user-stories with proposals that make it more comprehensive, while avoiding straying too far from the initial requirements?
\end{itemize}

%% file: sections/findings.tex
\section{\uppercase{Evaluation: Findings and discussion}}
\label{sec:findings}

In this section, we present and discuss the results obtained on twenty representative conversations with LLM, and then we discuss threats to validity.
\vspace{0.1cm}

\noindent
\textit{Step 1. Design options summary.}
At step 1 of each of the 20 conversations, LLM answers a list of high-level options (the design options summary). 
The design summaries contain from four to eight options and have an average of 6.3 options. 
To assess the stability and content relevance of these 20 summaries,
we developed a 2-prompt conversation 
launched five times. In the first prompt, we asked LLM to generate a report about the similarity of the 20 summaries. For evaluating this similarity, LLM has to identify common elements, based on identical names, synonyms, or terms with close semantics.
In each of the five launched analyses, we observed that from six to eight options appear in more than half of summaries, revealing that most of the summaries share quite all their options.
In the second prompt, we asked LLM to analyze more deeply the four most frequent options.
The two most frequent ones are  
\textit{User Management/Account Management} and 
\textit{Content Management}.
Then, these three following options appear in different orders: 
\textit{Interaction/browsing} or (exclusive) 
\textit{support/communication}
and always 
\textit{Subscription/Financial/Payment}.
This reveals a stability in the summaries built by LLM.
These elements support a positive answer to (Q1). 

\vspace{0.1cm}
\noindent
\textit{Step 2. List of user stories for the selected design options.}  
The result of this step is rather straightforward to deliver for LLM, as it consists of enumerating the user-stories corresponding to the selection of one or more design options, that it created itself by grouping user-stories. Nevertheless, Table \ref{table:QAimplications} shows that the number of user-stories grouped by LLM in a design option
(at step 1) varies from one conversation to another, even if they were conducted the same day (e.g. Conversations Id 16, 17, and 18). This mitigates the positive answer to (Q1), as this means, that, even if nearly similar options names are presented to the system designer, these options may correspond to different user story groups. 
\vspace{0.1cm}

\noindent
\textit{Step 3. Application of the implications computed by TCA.}
Tables \ref{table:QAimplications} and \ref{table:QAUS} report figures about the two results of step 3, i.e. applied implications and obtained user-stories respectively. 
Entrusting the task of applying the implications means that we are sufficiently confident in LLM ability to follow the application procedure described in the prompt and to enrich it.  
The set of binary implications we use has the property of being ``direct'' meaning that using the premises as input and applying the implications all at once provides all the user stories that can be inferred. This is an important property that eases LLM task.
In order to assess our confidence, we developed a rule engine ($RuleEng$\footnote{https://gite.lirmm.fr/gutierre/expeimplications}) that applies TCA implications whose premise appears in step 2 result. 
The output of the rule engine is the set of deduced user-stories.  
We then compare the implications applied by $RuleEng$ with those applied by LLM (Table \ref{table:QAimplications}), and the user-stories computed by $RuleEng$ with those provided by LLM (Table \ref{table:QAUS}).

Results show a difference in seven conversations between the implications applied by $RuleEng$ and those effectively applied by LLM (in boldface in Table \ref{table:QAimplications}). Among these seven conversations, in conversations Id 11 and 14, LLM applies some new implications in addition to those applied by $RuleEng$. 
In the other five cases, there is a significant difference between implications applied by $RuleEng$ and the ones applied by LLM, e.g conversation Id 4 for which only seven implications applied by $RuleEng$ (on 95) were used by LLM among ten. Identifying the cause of such behavior raises questions (e.g. misunderstanding of the prompt or larger use of knowledge).
The same evaluation was carried out for the computed user-stories (Table \ref{table:QAUS}). 
This table shows 6 conversations where numbers differs, that also present a difference between implications, with a similar trend, i.e. when fewer implications are applied, fewer user-stories are computed, and reversely. Regarding the conversation that presents a difference in implications and not in the user-stories (conversation Id 11), we suppose that LLM did not apply some implications it declared to have applied. 
This result gives us a relative confidence in the way that LLM applies the implications and derives the user-stories, and contributes to answer partly positively to (Q2).
We observe a significant number of conversations with low quality of implication application by LLM (about 1/3). A learned lesson is that, at this stage of LLM developement, after application of step 3 in real practice, it is recommended to compare the number of implications applied by both LLM and $RuleEng$. 
When the difference is significant, the designer can either discard the conversation, or try to redirect LLM.
\vspace{0.1cm}

\input{tables/QA_implications}

\input{tables/QA_userstories}

\noindent
\textit{Step 4. Upgrade of the user-stories using LLM.} 
Table \ref{table:QAcompletion} reports the user-stories improvements made by  LLM 
on the results from step 3.
In four conversations (Id 8, 15, 16, and 17), user-stories were removed from step 3, meaning that LLM 
possibly considered some being semantic duplicates in the list. Of these four conversations, only one (Id 17) presents differences in both Tables \ref{table:QAimplications} and \ref{table:QAUS}.
For all the conversations, we note that  LLM adds user-stories. The increase ranges between 2\% and 136\%, and is 38\% on average.
A human reviewing confirmed their added-value, while remaining in the expected scope of the website domain, that fully justifies the use of LLM. 
For these conversations, we can conclude positively to (Q3).

\input{tables/QA_completion}

\paragraph{Threats to validity}

Internal validity deals with datasets and tools quality. We refer the readers to the paper introducing the used dataset \cite{DBLP:journals/ijar/BazinGHMT24} which exposes the concerns related to its building. 

The uncontrolled element of this process is the LLM computation (e.g. summarizing), the fact that LLM parameters cannot be set in the current version we used, and 
knowledge it can bring. This corresponds to plausible current working condition for many  software designers. 
In order to assess   the ability of LLM to apply implications, 
we developed, apart from LLM, a rule engine named $RuleEng$ to systematically apply  implications and obtain the expected resulting user-stories. In addition, a systematic human review of LLM results ensured their coherency in relation with the task and input data (e.g. user-stories, implications). This systematic review also allowed to identify abnormal results, corresponding to a loss of quality of ChatGPT 4.o answers that has occurred during a short time, due to the change of its model. 
By nature of this tool, that shows randomness, we cannot have a perfect guarantee on the stability of the results and their repeatability.

We proposed various ways to assess the steps, i.e. a similarity study between the delivered design summaries for step 1, a comparison between the results of the rule engine and of  LLM for step 3, and checking whether updates are of reasonable size and do not fall outside the domain scope for step 4. 
Designing more in-depth assessments remains a task for the future.

The case study deserves to be extended in several directions before generalizing (external validity), using a richer user stories description included in \cite{DBLP:journals/ijar/BazinGHMT24}, and considering other SPL domains. 
Nevertheless the study allows to expect that the approach is relevant  
on datasets of the same size and nature (commercial and community websites). 
We also could have considered other LLMs, but the objective was not to determine whether one model is better than another, but rather to demonstrate the feasibility of using an LLM.

%% file: tables/QA_implications.tex
\begin{table}[htb]
\centering
\caption{Implications ($Implicat.$) applied to obtain the set of user-stories in step 3 per conversation. Selected Design options are expressed by their acronym (e.g. T stands for Transaction). $US$ stands for User-stories. 
Values in  bold face highlight the differences between implications applied by  LLM and the ones applied by RuleEng. 
%The lines correspond to the 20 conversations with chatGPT 4o. 
%The first four columns provide: the conversation identifier, its date, an acronym of the selected options (e.g. T stands for Transaction), and the number of initial user-stories $US$ corresponding to the selected options (output of step 2). 
%The next two columns show the number of implications that have to be applied (\#computed Imp.) compared to the number of implications the LLM indicates it applies (\#LLM answer Imp.). 
%The last column shows the number of common implications (\#common Imp.). 
%The number of implications that the LLM should apply, but does not apply is \#computed-\#common. The number of implications that the LLM applies but should not apply is \#LLM answer-\#common.
}
\label{table:QAimplications}
\adjustbox{width=\linewidth}{
\begin{tabular}{|c|c|c|c|c|c|c|}
\hline 
%\textbf{Convers.}&\textbf{\#Comput.}&\textbf{Options}&\textbf{\#initial US}&\textbf{\#comput. }&\textbf{\#LLM answer}&\textbf{\#common}\\
\textbf{Conver-}&\textbf{Compu-}&\textbf{Selected}&\textbf{\#initial}&\textbf{\#Implicat.}&\textbf{\#Implicat.}&\textbf{\#Implicat.}\\
\textbf{sation}&\textbf{tation}&\textbf{Options}&\textbf{US from}&\textbf{applied by}&\textbf{applied}&\textbf{applied by}\\
\textbf{Id}&\textbf{date}&\textbf{in step 2}&\textbf{step 2}&\textbf{RuleEng}&\textbf{by LLM}&\textbf{RuleEng and LLM}\\
%\textbf{Convers.}&\textbf{\#Comput.}&\textbf{Select.}&\textbf{\#initial}&\textbf{\#Applied }&\textbf{\#LLM answer}&\textbf{\#common}\\
%\textbf{Id}&\textbf{date}&\textbf{Options}&\textbf{US from}&\textbf{TCA}&\textbf{Implic.}&\textbf{Implic.}\\
%&&\textbf{at step2}&\textbf{step 2}&\textbf{Implic.}&\textbf{2}&\textbf{1}\\
\hline	
%T_T_1031_1 
1 & 10/31 & T &6&36&36&36\\\hline
%T_T_F_1031_2
2 & 10/31 & T/F &11&90&90&90\\\hline
%T_SP_1031_3
3 & 10/31 & SP &8&69&69&69\\\hline
%T\_SN\_SI\_1031\_4
4 & 10/31 & SN/SI &13&\textbf{95}&\textbf{10}&\textbf{7}\\\hline
%T\_PSM\_PM\_1031\_5
5 & 10/31 & PSM/PM &14&114&114&114\\\hline
%T\_PSM\_PM\_1031\_6
6 & 10/31 & PSM/PM &22&104&104&104\\\hline
%T\_FT\_1102\_1&&&\\\hline
7 & 11/02 & FT &6&35&35&35\\\hline
%T\_SPM\_1102\_2&&&\\\hline
8 & 11/02 & SPM &18&78&78&78\\\hline
%T\_FT\_1102\_3&&&\\\hline
9 & 11/02 & FT &7&\textbf{67}&\textbf{12}&\textbf{3}\\\hline
%T\_MPF\_1102\_4&&&\\\hline
10 &  11/02 & MPF &8&\textbf{48}&\textbf{4}&\textbf{3}\\\hline
%T\_FO\_1102\_5&&&\\\hline
11 & 11/02 & FO &5&\textbf{42}&\textbf{53}&\textbf{42}\\\hline
%T\_SPP\_1102\_6&&&\\\hline
12 & 11/02 & SPP &8&66&66&66\\\hline
%T\_PSM\_1102\_7&&&\\\hline
13 & 11/02  & PSM &11&\textbf{67}&\textbf{8}&\textbf{6}\\\hline
%T\_UIF\_1103\_1&&&\\\hline
14 & 11/03 & UIF &14&\textbf{46}&\textbf{112}&\textbf{46}\\\hline
%T\_PT\_1103\_2&&&\\\hline
15 & 11/03 & PT &8&64&64&64\\\hline
%T\_TM\_1103\_3&&&\\\hline
16 & 11/03 & TM &10&94&94&94\\\hline
%T\_TM\_1103\_4&&&\\\hline
17 & 11/03 & TM &10&\textbf{94}&\textbf{71}&\textbf{0}\\\hline
%T\_TM\_1103\_5&&&\\\hline
18 & 11/03 & TM &5&38&38&38\\\hline
%T\_TF\_1103\_6&&&\\\hline
19 & 11/03 & TF &9&65&65&65\\\hline
%T\_SPM\_1103\_7	&&&\\\hline
20 &  11/03  & SPM &6&63&63&63\\\hline
\end{tabular}
}%adjustbox
\end{table}

%% file: tables/QA_userstories.tex
\begin{table}[htb]
\centering
\caption{User-stories ($US$) computed at step 3 per conversation. 
%Selected Design options are expressed by their acronym (e.g. T stands for Transaction). 
Values in bold face highlight differences between US computed by LLM and those computed by RuleEng.
%\pma{je suppose que ce n'est pas TCA on plus ici. RuleEng ?}
%The lines corresponds to the 20 conversations with chatCPT4o. 
%The first four columns provide: the conversation  identifier, its date, an acronym of the selected options, and the number of user-stories corresponding to the selected options (output of step 2). 
%The next two columns show the number of user-stories that have been computed by an exact method (\#computed US) compared to the number of user-stories in the answer of the LLM at the end of step 3 (\#LLM answer US). 
%The last column shows the number of common user-stories (\#common US). 
%The number of user-stories that the LLM should have answered, but does not have answered is \#computed-\#common. The number of user-stories that the LLM answered but should not have answered is \#LLM answer-\#common.
}
\label{table:QAUS}
\adjustbox{width=\linewidth}{
\begin{tabular}{|c|c|c|c|c|c|c|}
\hline 
\textbf{Conver-}&\textbf{Compu-}&\textbf{Selected}&\textbf{\#initial}&\textbf{\#US }&\textbf{\#US}&\textbf{\#US}\\
\textbf{sation}&\textbf{tation}&\textbf{Options}&\textbf{US from}&\textbf{comput.}&\textbf{comput. }&\textbf{comput. By}\\
\textbf{Id}&\textbf{Date}&\textbf{in step 2}&\textbf{step 2}&\textbf{by RuleEng}&\textbf{by LLM}&\textbf{RuleEng and LLM}\\
\hline	
%T_T_1031_1 
1 & 10/31 & T &6&25&25&25\\\hline
%T_T_F_1031_2
2 & 10/31 & T/F &11&41&41&41\\\hline
%T_SP_1031_3
3 & 10/31 & SP &8&36&36&36\\\hline
%T\_SN\_SI\_1031\_4
4 & 10/31 & SN/SI &13&\textbf{40}&\textbf{14}&\textbf{14}\\\hline
%T\_PSM\_PM\_1031\_5
5 & 10/31 & PSM/PM &14&43&43&43\\\hline
%T\_PSM\_PM\_1031\_6
6 & 10/31 & PSM/PM &22&41&41&41\\\hline
%T\_FT\_1102\_1&&&\\\hline
7 & 11/02 & FT &6&25&25&25\\\hline
%T\_SPM\_1102\_2&&&\\\hline
8 & 11/02 & SPM &18&44&44&44\\\hline
%T\_FT\_1102\_3&&&\\\hline
9 & 11/02 & FT &7&\textbf{30}&\textbf{16}&\textbf{14}\\\hline
%T\_MPF\_1102\_4&&&\\\hline
10 &  11/02 & MPF &8&\textbf{27}&\textbf{11}&\textbf{8}\\\hline
%T\_FO\_1102\_5&&&\\\hline
11 & 11/02 & FO &5&31&31&31\\\hline
%T\_SPP\_1102\_6&&&\\\hline
12 & 11/02 & SPP &8&28&28&28\\\hline
%T\_PSM\_1102\_7&&&\\\hline
13 & 11/02  & PSM &11&\textbf{33}&\textbf{11}&\textbf{11}\\\hline
%T\_UIF\_1103\_1&&&\\\hline
14 & 11/03 & UIF &14&\textbf{27}&\textbf{50}&\textbf{27}\\\hline
%T\_PT\_1103\_2&&&\\\hline
15 & 11/03 & PT &8&28&28&28\\\hline
%T\_TM\_1103\_3&&&\\\hline
16 & 11/03 & TM &10&49&49&49\\\hline
%T\_TM\_1103\_4&&&\\\hline
17 & 11/03 & TM &10&\textbf{49}&\textbf{33}&\textbf{27}\\\hline
%T\_TM\_1103\_5&&&\\\hline
18 & 11/03 & TM &5&29&29&29\\\hline
%T\_TF\_1103\_6&&&\\\hline
19 & 11/03 & TF &9&28&28&28\\\hline
%T\_SPM\_1103\_7	&&&\\\hline
20 &  11/03  & SPM &6&31&31&31\\\hline
\end{tabular}
}%adjustbox
\end{table}

%% file: tables/QA_completion.tex
\begin{table}[htb]
\centering
\caption{User-stories ($US$) per conversation obtained in step 3 and 4. 
%Selected Design options are expressed by their acronyms (e.g. T stands for Transaction). 
Values in  bold face highlight differences between US obtained by  LLM in step 3 and those obtained in step 4.
%The lines corresponds to the 20 conversations with chatCPT4o. 
%The first four columns provide: the conversation  identifier, its date, an acronym of the selected options, and the number of user-stories corresponding to the selected options (output of step 2). 
%The next two columns show the number of user-stories after step 3 (\#Step 3 US) and after completion of step4 (\# Step 4 US). 
%The last column shows the number of common user-stories (\#common US), to check if the LLM preserved all the user-stories of step 3, or removed some of them.
}
\label{table:QAcompletion}
\adjustbox{width=\linewidth}{
\begin{tabular}{|c|c|c|c|c|c|c|}
\hline 
%\textbf{Id}&\textbf{Date}&\textbf{Options step2}&\textbf{\#initial US}&\textbf{\#step 3  US}&\textbf{\#step 4 US}&\textbf{\#common US}\\
\textbf{Conver-}&\textbf{Compu-}&\textbf{Selected}&\textbf{\#initial}&\textbf{\#US }&\textbf{\#US}&\textbf{\#US listed}\\
\textbf{sation}&\textbf{tation}&\textbf{Options}&\textbf{US from}&\textbf{listed}&\textbf{listed}&\textbf{both in}\\
\textbf{Id}&\textbf{Date}&\textbf{in step 2}&\textbf{step 2}&\textbf{in step 3}&\textbf{in step 4}&\textbf{step 3 and 4}\\

\hline	
%T_T_1031_1 
1 & 10/31 & T &6&25&37&25\\\hline
%T_T_F_1031_2
2 & 10/31 & T/F &11&41&55&41\\\hline
%T_SP_1031_3
3 & 10/31 & SP &8&36&45&36\\\hline
%T\_SN\_SI\_1031\_4
4 & 10/31 & SN/SI &13&14&23&14\\\hline
%T\_PSM\_PM\_1031\_5
5 & 10/31 & PSM/PM &14&43&52&43\\\hline
%T\_PSM\_PM\_1031\_6
6 & 10/31 & PSM/PM &22&41&53&41\\\hline
%T\_FT\_1102\_1&&&\\\hline
7 & 11/02 & FT &6&25&33&25\\\hline
%T\_SPM\_1102\_2&&&\\\hline
8 & 11/02 & SPM &18&\textbf{44}&\textbf{52}&\textbf{40}\\\hline
%T\_FT\_1102\_3&&&\\\hline
9 & 11/02 & FT &7&16&24&16\\\hline
%T\_MPF\_1102\_4&&&\\\hline
10 &  11/02 & MPF &8&11&18&11\\\hline
%T\_FO\_1102\_5&&&\\\hline
11 & 11/02 & FO &5&31&38&31\\\hline
%T\_SPP\_1102\_6&&&\\\hline
12 & 11/02 & SPP &8&28&35&28\\\hline
%T\_PSM\_1102\_7&&&\\\hline
13 & 11/02  & PSM &11&11&17&11\\\hline
%T\_UIF\_1103\_1&&&\\\hline
14 & 11/03 & UIF &14&50&57&50\\\hline
%T\_PT\_1103\_2&&&\\\hline
15 & 11/03 & PT &8&\textbf{28}&\textbf{35}&\textbf{24}\\\hline
%T\_TM\_1103\_3&&&\\\hline
16 & 11/03 & TM &10&\textbf{49}&\textbf{50}&\textbf{44}\\\hline
%T\_TM\_1103\_4&&&\\\hline
17 & 11/03 & TM &10&\textbf{33}&\textbf{78}&\textcolor{black}{32}\\\hline
%T\_TM\_1103\_5&&&\\\hline
18 & 11/03 & TM &5&29&38&29\\\hline
%T\_TF\_1103\_6&&&\\\hline
19 & 11/03 & TF &9&28&37&28\\\hline
%T\_SPM\_1103\_7	&&&\\\hline
20 &  11/03  & SPM &6&31&42&31\\\hline
\end{tabular}
}%adjustbox
\end{table}

%% file: sections/relatedWork.tex
\section{\uppercase{Related Work}}
\label{sec:relatedwork}

LLMs provide many opportunities for achieving software engineering tasks, as it has been reported in 
a recent systematic literature review \cite{10.1145/3695988}. 
Two works at the requirement stage are worth mentioning. 
An approach for synthesizing  specifications of software configurations from natural language texts has been proposed in \cite{DBLP:journals/corr/abs-2304-09181}. Here we do not rely on identifying specifications, as we dispose of user-stories, which are formatted expressions of specifications. 
LLM is used to evaluate the quality of user-stories in
\cite{10.1007/978-3-031-48550-3_17}. In our present work, we do not evaluate the user-story sets and we consider they have a sufficient quality level to serve as a reference basis for building a new user-story set.
A comparison of two approaches (rules versus LLM) to derive UML sequence diagrams from user stories is presented in \cite{10.1145/3640310.3674081}. 
Here, we do not aim to derive diagrammatic representations.

Domain models have been derived from user-stories using approaches including LLM interaction in \cite{DBLP:conf/models/ArulmohanM023,DBLP:conf/re/BragilovskiCDS24}.
In \cite{DBLP:conf/re/BragilovskiCDS24}, examples of extracted domain concepts are personas, actions or entities. They used the reference dataset in \cite{dalpiaz2018ds}, which contains user-story sets for single systems on different topics,  and has been introduced in \cite{DBLP:journals/infsof/DalpiazSBAL19}.
In Step 1, we do not extract a domain model, rather we ask LLM to categorize the roles and features, thus to operate on this domain model to give a synthetic view of high-level design options. The dataset we use  contains a family of user-story sets.  

To our knowledge, there are few works that integrate SPL and LLM. 
One direction consists in applying Software Product Line Engineering (SPLE) principles to construct composite LLMs \cite{10.1145/3646548.3676599}.
In another direction, LLMs are used to achieve or assist with certain tasks of the SPLE life cycle, as we do in this paper. 
E.g. ChatGPT was used to synthesize SPL on the basis of a set of variants in \cite{DBLP:conf/splc/AcherM23}. 
In this latter paper, different types of system variants are considered: Java, UML, GraphML, state charts, and PNG.
We follow this line of research with a few differences.
Variability is identified using an exact method (i.e. TCA).
When asking LLM to identify design options that group roles and features, the design options are a way to annotate the user-stories, which can be considered as a part of the product line to a certain extent. 
As suggested in the discussion in \cite{DBLP:conf/splc/AcherM23}, our proposal combines the use of LLM with a deterministic approach.

%% file: sections/conclusion.tex
\section{\uppercase{Conclusion}}
\label{sec:conclusion}

In this paper, we investigated the combination of LLM with a logical analysis method (TCA), applied to a user-story sets family in order to assist software engineers in the building of a new user-story set.
 The method uses (1) the knowledge extracted from the user-story sets family to frame the scope and guide towards valid configurations, 
and (2) 
knowledge of LLM to overcome the limitations inherent to the existing system family. 
%It has been applied to a dataset from the literature.

This work can be extended in several directions.
First, TCA provides additional kinds of implications, not considered in this study, from which other types of logical dependencies (e.g. mutual exclusions) can be inferred. They can be used to fine-tune the software's final configuration.
To address higher dimensions, like the purpose or the version, Polyadic Concept Analysis \cite{DBLP:journals/order/Voutsadakis02} can be used. %
Second, the process can be refined to better match designers' needs.
For instance, 
 LLM can propose various abstraction level options, 
or 
the implications provided by the rule engine can be used without requiring  LLM to apply them. 
This may reduce the sensitivity of the configuration to the randomness of  LLM.
%The LLM could also be used to  enumerate various possible combinations (i.e. sampling), based on logical dependencies and knowledge to let the software engineers choosing which one they want.

%% file: main_enase.bbl
\begin{thebibliography}{}

\bibitem[Acher and Martinez, 2023]{DBLP:conf/splc/AcherM23}
Acher, M. and Martinez, J. (2023).
\newblock Generative {AI} for reengineering variants into software product
  lines: An experience report.
\newblock In {\em Proc. of the 27th {ACM} Int. Systems and Software Product
  Line Conf. - Volume B, {SPLC} 2023}, pages 57--66. {ACM}.

\bibitem[Arulmohan et~al., 2023]{DBLP:conf/models/ArulmohanM023}
Arulmohan, S., Meurs, M., and Mosser, S. (2023).
\newblock Extracting domain models from textual requirements in the era of
  large language models.
\newblock In {\em 5th Ws. on Artificial Intelligence and Model-driven Eng. @
  {ACM/IEEE} {MODELS} 2023}, pages 580--587. {IEEE}.

\bibitem[Bazin et~al., 2024]{DBLP:journals/ijar/BazinGHMT24}
Bazin, A., Georges, T., Huchard, M., Martin, P., and Tibermacine, C. (2024).
\newblock Exploring the 3-dimensional variability of websites' user-stories
  using triadic concept analysis.
\newblock {\em Int. J. Approx. Reason.}, 173:109248.

\bibitem[Bragilovski et~al., 2024]{DBLP:conf/re/BragilovskiCDS24}
Bragilovski, M., van Can, A.~T., Dalpiaz, F., and Sturm, A. (2024).
\newblock Deriving domain models from user stories: Human vs. machines.
\newblock In {\em 32nd {IEEE} Int. Requirements Engineering Conf., {RE} 2024},
  pages 31--42. {IEEE}.

\bibitem[Dalpiaz, 2018]{dalpiaz2018ds}
Dalpiaz, F. (2018).
\newblock Requirements data sets (user stories).
\newblock {Mendeley Data, V1, doi: 10.17632/7zbk8zsd8y.1}.

\bibitem[Dalpiaz et~al., 2019]{DBLP:journals/infsof/DalpiazSBAL19}
Dalpiaz, F., Schalk, I. V.~D., Brinkkemper, S., Aydemir, F.~B., and Lucassen,
  G. (2019).
\newblock Detecting terminological ambiguity in user stories: Tool and
  experimentation.
\newblock {\em Inf. Softw. Technol.}, 110:3--16.

\bibitem[Ganter and Obiedkov, 2004]{DBLP:conf/iccs/GanterO04}
Ganter, B. and Obiedkov, S.~A. (2004).
\newblock Implications in triadic formal contexts.
\newblock In {\em Conceptual Structures at Work: 12th {ICCS} 2004}, volume 3127
  of {\em LNCS}, pages 186--195. Springer.

\bibitem[Ganter and Wille, 1999]{DBLP:books/daglib/0095956}
Ganter, B. and Wille, R. (1999).
\newblock {\em Formal Concept Analysis - Mathematical Foundations}.
\newblock Springer.

\bibitem[Gomez-Vazquez and Cabot, 2024]{10.1145/3646548.3676599}
Gomez-Vazquez, M. and Cabot, J. (2024).
\newblock Exploring the use of software product lines for the combination of
  machine learning models.
\newblock In {\em Proc. of the 28th ACM Int. Systems and Software Product Line
  Conference}, SPLC '24, page 26–29. ACM.

\bibitem[Hou et~al., 2024]{10.1145/3695988}
Hou, X., Zhao, Y., Liu, Y., Yang, Z., Wang, K., Li, L., Luo, X., Lo, D.,
  Grundy, J., and Wang, H. (2024).
\newblock Large language models for software engineering: A systematic
  literature review.
\newblock {\em ACM Trans. Softw. Eng. Methodol.}, 33(8).

\bibitem[Jahan et~al., 2024]{10.1145/3640310.3674081}
Jahan, M., Hassan, M.~M., Golpayegani, R., Ranjbaran, G., Roy, C., Roy, B., and
  Schneider, K. (2024).
\newblock {Automated Derivation of UML Sequence Diagrams from User Stories:
  Unleashing the Power of Generative AI vs. a Rule-Based Approach}.
\newblock In {\em Proc. of the ACM/IEEE 27th Int. Conf. on Model Driven
  Engineering Languages and Systems}, MODELS '24, page 138–148. ACM.

\bibitem[Lehmann and Wille, 1995]{DBLP:conf/iccs/LehmannW95}
Lehmann, F. and Wille, R. (1995).
\newblock A triadic approach to formal concept analysis.
\newblock In {\em 3rd Int. Conf. on Conceptual Structures, {ICCS} '95}, volume
  954 of {\em LNCS}, pages 32--43. Springer.

\bibitem[Lucassen et~al., 2016]{UsAsRequirment}
Lucassen, G., Dalpiaz, F., Van~der Werf, J.~M., and Brinkkemper, S. (2016).
\newblock Improving agile requirements: the quality user story framework and
  tool.
\newblock {\em Requirements Engineering}, 21.

\bibitem[Mandal et~al., 2023]{DBLP:journals/corr/abs-2304-09181}
Mandal, S., Chethan, A., Janfaza, V., Mahmud, S. M.~F., Anderson, T.~A., Turek,
  J., Tithi, J.~J., and Muzahid, A. (2023).
\newblock Large language models based automatic synthesis of software
  specifications.
\newblock {\em CoRR}, abs/2304.09181.

\bibitem[Mondal et~al., 2024]{DBLP:conf/msr/MondalBR24}
Mondal, S., Bappon, S.~D., and Roy, C.~K. (2024).
\newblock Enhancing user interaction in chatgpt: Characterizing and
  consolidating multiple prompts for issue resolution.
\newblock In {\em 21st {IEEE/ACM} Int. Conf. on Mining Software Repositories},
  pages 222--226. {ACM}.

\bibitem[Pohl et~al., 2005]{DBLPbooks/daglib/0015277}
Pohl, K., B{\"{o}}ckle, G., and van~der Linden, F. (2005).
\newblock {\em Software Product Line Engineering - Foundations, Principles, and
  Techniques}.
\newblock Springer.

\bibitem[Ronanki et~al., 2024]{10.1007/978-3-031-48550-3_17}
Ronanki, K., Cabrero-Daniel, B., and Berger, C. (2024).
\newblock Chatgpt as a tool for user story quality evaluation: Trustworthy
  out of the box?
\newblock In {\em Agile Processes in Software Engineering and Extreme
  Programming -- Workshops}, pages 173--181, Cham. Springer Nature Switzerland.

\bibitem[Schulhoff et~al., 2024]{schulhoff2024promptreportsystematicsurvey}
Schulhoff, S., Ilie, M., Balepur, N., Kahadze, K., Liu, A., Si, C., Li, Y.,
  Gupta, A., Han, H., Schulhoff, S., Dulepet, P.~S., Vidyadhara, S., Ki, D.,
  Agrawal, S., Pham, C., Kroiz, G., Li, F., Tao, H., Srivastava, A., Costa,
  H.~D., Gupta, S., Rogers, M.~L., Goncearenco, I., Sarli, G., Galynker, I.,
  Peskoff, D., Carpuat, M., White, J., Anadkat, S., Hoyle, A., and Resnik, P.
  (2024).
\newblock The prompt report: A systematic survey of prompting techniques.
\newblock https://arxiv.org/abs/2406.06608.

\bibitem[Voutsadakis, 2002]{DBLP:journals/order/Voutsadakis02}
Voutsadakis, G. (2002).
\newblock Polyadic concept analysis.
\newblock {\em Order}, 19(3):295--304.

\bibitem[White et~al., 2023]{DBLP:journals/corr/abs-2302-11382}
White, J., Fu, Q., Hays, S., Sandborn, M., Olea, C., Gilbert, H., Elnashar, A.,
  Spencer{-}Smith, J., and Schmidt, D.~C. (2023).
\newblock {A Prompt Pattern Catalog to Enhance Prompt Engineering with
  ChatGPT}.
\newblock {\em CoRR}, abs/2302.11382.

\end{thebibliography}
